\documentclass[letterpaper, aps, showpacs, onecolumn]{revtex4}

\usepackage{amsmath}
\usepackage{amssymb}
\usepackage{latexsym}
\usepackage{graphicx,epstopdf}
\usepackage{hyperref}

\begin{document}

\title{On Newton's equation of motion with friction and stochastic noise,\\
the Ostrogradsky-instability and the hierarchy of environments\\
An application of the Onsager-Machlup theory II}
\author{Alexander Jurisch}
\affiliation{ajurisch@ymail.com, Munich, Germany}

\begin{abstract}
Onsager and Machlup proposed a second order variational-principle in order to include inertial effects into the Langevin-equation, giving a Lagrangian with second order derivatives in time. This but violates Ostrogradysky's theorem, which proves that Lagrangians with higher than first order derivatives are meaningless. As a consequence, inertial effects cannot be included in a standard way. By using the canonical formalism, we suggest a solution to this fundamental problem. Furthermore, we provide elementary arguments about the hierarchy of immersions and actions between an ideal system and several environments and show, that the structure of the Lagrangian sensitively depends on this hierarchy.
\end{abstract}
\maketitle
\section{Introduction}
This sequel of our paper on the Onsager-Machlup theory \cite{Jurisch1} is inspired by a serious mistake we noticed in a paper by Taniguchi and Cohen \cite{Taniguchi}. As it turned out, this mistake has already been made by Onsager and Machlup in their seminal paper \cite{Machlup}, and has been applied ever since. In the attempt to include inertial effects into the Langevin-equation Onsager and Machlup proposed a second order variational-principle based on a generalization of the Onsager-Machlup Lagrangian. This but violates Ostrogradsky's theorem \cite{Ostrogradsky}. In the following we elucidate the source of the mistake and suggest a solution of the problem.

The equation of motion of interest is given by
\begin{equation}
m\,\ddot{q}(t)\,=\,F[q(t)]\,+\,\mu_{1}\,\dot{q}(t)\,+\,\sqrt{\sigma_{2}}\,\eta(t)\quad,
\label{Introduction1}\end{equation}
where $F[q(t)]$ is a conservative force-field, the second term describes Stokes-friction, and the last term is a Gaussian stochastic process. Thus, it is assumed that already the acceleration is distorted by noise. In \cite{Taniguchi} as in \cite{Machlup} friction has been introduced by purely phenomenological arguments.

By the assumption that the motion takes place in the over-damped regime, $\ddot{q}(t)\,=\,0$, Eq. (\ref{Introduction1}) can be cast into
\begin{equation}
\dot{q}(t)\,=\,-\,\mu_{1}^{-1}\,\left(F[q(t)]\,+\,\sqrt{\sigma_{2}}\,\eta(t)\right)\quad,
\label{Introduction2}\end{equation}
which may be identified as a Langevin-equation, from which the corresponding Onsager-Machlup Lagrangian can be derived. We remark that such a reinterpretation already becomes impossible when Newtonian friction $\sim\,\dot{q}^{2}$ is present, see \cite{Jurisch1}.

In a next step, following Onsager and Machlup \cite{Machlup}, the acceleration is included as an additional drift-field
\begin{equation}
\dot{q}(t)\,=\,\frac{m}{\mu_{1}}\,\ddot{q}(t)\,-\,\mu_{1}^{-1}\,\left(F[q(t)]\,+\,\sqrt{\sigma_{2}}\,\eta(t)\right)\quad,
\label{Introduction3}\end{equation}
which is just a rearrangement of Eq. (\ref{Introduction1}). Along with \cite{Machlup} Eq. (\ref{Introduction3}) is still interpreted as a Langevin-equation, which is a highly doubtful interpretation. Furthermore, again following Onsager and Machlup \cite{Machlup}, the Onsager-Machlup Lagrangian is set up by
\begin{equation}
\mathcal{L}(\ddot{q},\,\dot{q},\,q)\,=\,\frac{\mu_{1}^{2}}{2\,\sigma_{2}}\,\left(\dot{q}(t)\,-\,\frac{m}{\mu_{1}}\,\ddot{q}(t)\,+\,\mu_{1}^{-1}\,F[q(t)]\right)^{2}\quad.
\label{Introduction4}\end{equation}
However, this Lagrangian is completely meaningless. The fundamental, but unfortunately scarcely known theorem of Ostrogradsky \cite{Ostrogradsky} proves that Lagrangians, which depend on higher than first order derivatives in a non-degenerate way do not describe any meaningful dynamics. Non-degeneracy here means that $\partial\mathcal{L}/\partial\ddot{q}$ is a function of $\ddot{q}$, to which the Lagrangian Eq. (\ref{Introduction4}) applies for. Ostrogradsky's theorem holds in general, it is one of the two hard criteria every Lagrangian, which claims physical content must obey. The second hard criterion are the Helmholtz-conditions \cite{Helmholtz}, see also Nigam and Banerjee \cite{Nigam},  Nucci and Leach \cite{Nucci}, and references therein. A violation of one of these criteria is sufficient to render a Lagrangian meaningless. Consequently, the Lagrangian under consideration here does not exist. This clarifies that the interpretation of Eq. (\ref{Introduction3}) as a Langevin-equation is ineligible. Furthermore, this rules out the second order variational-principle given by Onsager and Machlup in \cite{Machlup}. As a consequence, inertial effects cannot be included in a standard way.  We have not checked if the Lagrangian Eq. (\ref{Introduction4}) also violates the Helmholtz-conditions, since the violation of Ostrogradsky's theorem is sufficient to rule this ansatz out.

Systems described by Lagrangians with higher than first order derivatives suffer from the Ostrogradsky-instability. In short, the phase-space of such systems either explodes or decays almost instantly. This comes from pathological excitations and states of unbounded positive and negative energy. We emphasize that this must not be confused with the Dirac-theory. The Dirac-Lagrangian is sound with respect to it's derivatives, the spinors are relativistic fields and the unbounded scale of energy has a precise physical meaning.

To illustrate the origin of the Ostrogradsky-instability, we calculate the Hamiltonian of the Lagrangian Eq. (\ref{Introduction4}). The canonic coordinates are given by
\begin{equation}
Q_{1}\,=\,q,\quad Q_{2}\,=\,\dot{q},\quad P_{1}\,=\,\frac{\partial\,\mathcal{L}}{\partial\,\dot{q}}\,-\,\frac{d}{d\,t}\,\frac{\partial\,\mathcal{L}}{\partial\,\ddot{q}},\quad P_{2}\,=\,\frac{\partial\,\mathcal{L}}{\partial\,\ddot{q}}\quad.
\label{Introduction5}\end{equation}
It is easy to see that this choice is the natural generalization of the first order case. For our present system, this gives
\begin{equation}
P_{2}\,=\,\frac{m}{\sigma_{2}}\,\left(m\,\ddot{q}(t)\,-\,\mu_{1}\,\dot{q}(t)\,-\,F[q(t)]\right)\,\rightarrow\,\ddot{q}(t)\,=\,\mathcal{Q}(P_{2},\,Q_{1},\,Q_{2})\,=\,\frac{\sigma_{2}}{m^{2}}\,P_{2}\,+\,\frac{\mu_{1}}{m}\,Q_{2}\,+\,m^{-1}\,F[Q_{1}]\quad.
\end{equation}
The Hamiltonian follows on the standard route
\begin{eqnarray}
\mathcal{H}(P_{2},\,P_{1},\,Q_{2},\,Q_{1})&=&P_{1}\,Q_{2}\,+\,P_{2}\,\mathcal{Q}(P_{2},\,Q_{1},\,Q_{2})\,-\,\mathcal{L}\left(\mathcal{Q}(P_{2},\,Q_{1},\,Q_{2}),\,Q_{2},\,Q_{1}\right)\nonumber\\
&=&P_{1}\,Q_{2}\,+\,\frac{\sigma_{2}}{2\,m^{2}}\,P_{2}^{2}\,+\,P_{2}\,\left(\frac{\mu_{1}}{m}\,Q_{2}\,+\,m^{-1}\,F[Q_{1}]\right)\quad.
\label{Introduction6}\end{eqnarray}
The term linear in $P_{2}$ goes without problems, it can be removed by quadratic completion and acts like a gyroscopic potential. The term linear in $P_{1}$ but is non-trivial, since it does not resolve by Legendre-transform. The momentum $P_{1}$ is not subject to any constraints, it is completely unbounded and can take any value. By coupling to $Q_{2}$ this lack of boundary tends to disrupt the whole system. It is now easy to see that still higher order derivatives in the Lagrangian make things only worse. This is the origin of the Ostrogradsky-instability. By calculating the canonic equations and going back to the Euler-Lagrange equations one can convince oneself that Ostrogradsky's choice for the canonic coordinates indeed is correct.

For the whole impact of Ostrogradsky's theorem, please see e.g. Motomashi \cite{Motomashi}, especially Woodard \cite{Woodard}, and references therein. For completeness we add that Eq. (\ref{Introduction4}) must not be confused with the Gaussian action-principle, see e.g. Lanczos \cite{Lanczos}. The Gaussian action is no Lagrangian, but an application of the method of least squares.
\newline

\emph{Short historical note}: Because Ostrogradsky's theorem is unbeknownst to a broader audience, we think some words about it's history are in order. Ostrogradsky discovered his theorem in 1850, but it got almost unnoticed. At this time the understanding of energy and stability was still in it's childhood, and nobody could grasp what Ostrogradsky's theorem really says. Obviously even Landau and Lifshitz have not known about Ostrogradsky's theorem, nothing else could explain why it is not to be found in their textbook about mechanics. Thus, there is no wonder about that also Onsager and Machlup have not been aware of Ostrogradsky's theorem.

Ostrogradsky's theorem was rediscovered in particle physics, general relativity and cosmology. As far as we know, the first who proposed higher order Lagrangians were Pais and Uhlenbeck \cite{Pais}, ironically in 1950. Skeptics asked if this is allowed. On this route Ostrogradsky's theorem was rediscovered and is a nuisance in particle physics and cosmology ever since.

\section{Suggested solution of the problem}
In this section we shall elucidate our suggestion for a solution of the problem posed by Eqs. (\ref{Introduction1}). Since the Newtonian equation of motion with noise is not tractable by the Onsager-Machlup theory, we need two first-order equations with independent variables to introduce Langevin-equations. Thus, we employ the canonical formalism.

The equation of motion with Stokes-friction
\begin{equation}
\ddot{q}(t)\,=\,-\,\frac{1}{m}\,\partial_{q}U_{\rm{I}}[q(t)]\,+\,\mu_{1}\,\dot{q}(t)\quad,
\label{Solution7}\end{equation}
is a consequence of the Onsager-Machlup Lagrangian, see \cite{Jurisch1},
\begin{equation}
\mathcal{L}(\dot{q},\,q)\,=\,b(t)^{-2}\,\left(\frac{m}{2}\,\dot{q}(t)^{2}\,-\,U_{\rm{I}}[q(t)]\right)\,=\,\exp[-\,\mu_{1}\,t]\,\left(\frac{m}{2}\,\dot{q}(t)^{2}\,-\,U_{\rm{I}}[q(t)]\right)\quad.
\label{Solution8}\end{equation}
The function $b(t)=\exp[\mu_{1}/2\,t]$ is a Helmholtz-factor or, likewise, a Jacobi-multiplier. The Hamiltonian thus reads
\begin{equation}
\mathcal{H}(p,\,q)\,=\,\exp[\mu_{1}\,t]\,\frac{p^{2}(t)}{2\,m}\,+\,\exp[-\,\mu_{1}\,t]\,U_{\rm{I}}[q(t)]\quad,
\label{Solution9}\end{equation}
which also is known as the Caldirola-Kanai Hamiltonian \cite{Caldirola, Kanai}.
The canonic equations follow by
\begin{eqnarray}
dq(t)&=&\partial_{p}\mathcal{H}(p,\,q)\,=\,\exp[\mu_{1}\,t]\,\frac{p(t)}{m}\,dt\quad,\nonumber\\
dp(t)&=&-\,\partial_{q}\mathcal{H}(p,\,q)\,=\,-\,\exp[-\,\mu_{1}\,t]\,\partial_{q}U_{\rm{I}}[q(t)]\,dt\,+\,\sqrt{\sigma_{2}}\,dW(t)\quad,
\label{Solution10}\end{eqnarray}
where we already have introduced the stochastic momentum-process with variance $\sigma_{2}$. The Onsager-Machlup Lagrangian of the momentum-process then is
\begin{equation}
\mathcal{L}(\dot{p},\,q)\,=\,\frac{1}{2\,\sigma_{2}}\,\left(\dot{p}(t)\,+\,\exp[-\,\mu_{1}\,t]\,\partial_{q}U_{\rm{I}}[q(t)]\right)^{2}\quad.
\label{Solution11}\end{equation}
The new Hamiltonian of the system can now safely be written by
\begin{equation}
\mathcal{H}(p,\,P,\,q)\,=\,\exp[\mu_{1}\,t]\,\frac{p(t)^{2}}{2\,m}\,+\,\frac{\sigma_{2}}{2}\,P(t)^{2}\,-\,\exp[-\,\mu_{1}\,t]\,\partial_{q}U_{\rm{I}}[q(t)]\,P(t)\quad,
\label{Solution13}\end{equation}
where the second and third terms follow by a Legendre-transform of the Lagrangian Eq. (\ref{Solution11}. The unified canonic equations finally yield
\begin{eqnarray}
\dot{q}(t)&=&\partial_{p}\mathcal{H}(p,\,P,\,q)\,=\,\exp[\mu_{1}\,t]\,\frac{p(t)}{m}\quad,\nonumber\\
\dot{p}(t)&=&\partial_{P}\mathcal{H}(p,\,P,\,q)\,=\,\sigma_{2}\,P(t)\,-\,\exp[-\,\mu_{1}\,t]\,\partial_{q}U_{\rm{I}}[q(t)]\quad,\nonumber\\
\dot{P}(t)&=&-\,\partial_{q}\mathcal{H}(p,\,P,\,q)\,=\,\exp[-\,\mu_{1}\,t]\,\partial_{q}^{2}U_{\rm{I}}[q(t)]\,P(t)\quad.
\label{Solution14}\end{eqnarray}
For $\sigma_{2}=0$ the canonic equations decouple, and we recover Eq. (\ref{Solution10}), as it must. As a last step, we eliminate the momenta and calculate the Newtonian equation of motion for the the most probable path $q(t)$. The calculation is easily done by differentiating the first line in Eq. (\ref{Solution14}) and insert the other equations. This yields
\begin{equation}
\ddot{q}(t)\,=\,-\,\frac{1}{m}\,\partial_{q}U_{\rm{I}}[q(t)]\,+\,\mu_{1}\,\dot{q}(t)\,+\,\frac{\sigma_{2}}{m}\,\exp[\mu_{1}\,t]\,\exp\left[\int_{t_{0}}^{t}d\tau\,\exp[-\,\mu_{1}\,\tau]\,\partial_{q}^{2}U_{\rm{I}}[q(\tau)]\right]\quad.
\label{Solution15}\end{equation}
This Newtonian equation of motion elucidates that the noisy distortion influences the trajectory of the most probable path as additional force, which but depends on the interaction $U_{\rm{I}}[q(t)]$. Note that this force still remains finite for $\partial_{q}^{2}U_{\rm{I}}[q(\tau)]\,=\,0$. Furthermore, the additional force grows or decays as time evoles. If the force grows, this means that the stochastic noise tends to destroy the action of the potential and the friction, while a decay leaves the action of the potential and the friction sound or even enhances it.

We can integrate Eq. (\ref{Solution15}) and obtain the Onsager-Machlup Lagrangian for the combined system, reading
\begin{equation}
\mathcal{L}(\dot{q},\,q)\,=\,\exp[-\,\mu_{1}\,t]\,\left(\frac{m}{2}\,\dot{q}(t)^{2}\,-\,U_{\rm{I}}[q(t)]\right)\,+\,\sigma_{2}\,q(t)\,\exp\left[\int_{t_{0}}^{t}d\tau\,\exp[-\,\mu_{1}\,\tau]\,\partial_{q}^{2}U_{\rm{I}}[q(\tau)]\right]\quad.
\label{Solution16}\end{equation}
In principal, it is always possible to derive the Newtonian equation of motion from the canonic equations, but a reconstruction of the primal Lagrangian of the system may not always be possible. We chose the word \emph{primal} here to emphasize that this Lagrangian combines the environmental effects all in one, and thus is the Lagrangian to start with for all further examinations. We remark that the time-dependent exponential, last term in Eq. (\ref{Solution16}), shows the general structure of a Helmholtz-factor.

\section{Notes on the hierarchy of immersions and actions of environments}
Following our classification of interactions in \cite{Jurisch1}, we see that the additional potential in Eq. (\ref{Solution16}) acts like an external potential $U_{\rm{E}}[q(t),\,t]$. The first term in the Lagrangian Eq. (\ref{Solution16}) describes an ideal harmonic oscillator, that is immersed in environment (I), which creates Stokes-friction. The stochastic process disturbs the environment (I) from the outside, and consequently may be understood as environment (II), that acts upon environment (I). This leads to a hierarchy, that can be written by sets
\begin{equation}
\{\rm{harmonic\,oscillator}\,\subset\,\rm{environment (I)}\}\,\leftarrow\,\rm{environment (II)}\quad.
\label{Solution17}\end{equation}
We chose to write $\subset$ instead of $\in$, since we understand the ideal system not as to be an element of environment (I), but as a system immersed into environment (I). Thus, it is an additional subset of environment (I), which can be taken out again without reducing environment (I). If the ideal system would be an element, then a take-out would reduce environment (I).

Some more remarks have to be made about the system described in \cite{Taniguchi}. There, the potential $U[q(t)]$ is used as an external harmonic potential, that traps Brownian particles. To our regards, the correct initial ansatz for this system must be written by
\begin{equation}
\mathcal{L}(\dot{q},\,q)\,=\,\exp[-\,\mu_{1}\,t]\,\frac{m}{2}\,\dot{q}(t)^{2}\quad,
\label{Solution18}\end{equation}
since the harmonic trap acts from the outside. The Lagrangian Eq. (\ref{Solution18}) generates a Newtonian equation of motion with Stokes-friction. The system of Brownian particles itself is thus solely described by the kinetic energy plus the noise, while the harmonic trap surrounds this system, and consequently acts from the outside. According to our results in \cite{Jurisch1}, this modifies the equation of motion, and thus also leads to a different primal Lagrangian. In this interpretation, the hierarchy of systems as described in \cite{Taniguchi} is given by
\begin{equation}
\left\{\{\rm{free\,particles}\,\subset\,\rm{environment (I)}\}\,\leftarrow\,\rm{environment (II)}\right\}\,\leftarrow\,\rm{harmonic\,trap}\quad.
\label{Solution19}\end{equation}
This also means that the external harmonic potential $U_{\rm{E}}[q(t)]$ must only be added to the primal Onsager-Machlup Lagrangian as an external potential. This hierarchy then leads to
\begin{equation}
\mathcal{L}(\dot{q},\,q)\,=\,\left\{\left\{\exp[-\,\mu_{1}\,t]\,\frac{m}{2}\,\dot{q}(t)^{2}\right\}\,+\,\sigma_{2}\,q(t)\right\}\,-\,U_{\rm{E}}[q(t)]\quad.
\label{Solution20}\end{equation}

If, and only if environment (II) shall also disturb the external harmonic trap, then the initial Lagrangian can be taken by
\begin{equation}
\mathcal{L}(\dot{q},\,q)\,=\,\exp[-\,\mu_{1}\,t]\,\frac{m}{2}\,\dot{q}(t)^{2}\,-\,U_{\rm{E}}[q(t)]\quad.
\label{Solution21}\end{equation}
The set-relation in this case yields
\begin{equation}
\left\{\{\rm{free\,particles}\,\subset\,\rm{environment (I)}\}\,\leftarrow\,\rm{harmonic\,trap}\right\}\,\leftarrow\,\rm{environment (II)}\quad,
\label{Solution22}\end{equation}
and the primal Lagrangian follows by
\begin{equation}
\mathcal{L}(\dot{q},\,q)\,=\,\left\{\left\{\exp[-\,\mu_{1}\,t]\,\frac{m}{2}\,\dot{q}(t)^{2}\right\}\,-\,U_{\rm{E}}[q(t)]\right\}\,+\,\sigma_{2}\,q(t)\,\exp\left[\int_{t_{0}}^{t}d\tau\,\,\partial_{q}^{2}U_{\rm{E}}[q(\tau)]\right]\quad.
\label{Solution23}\end{equation}

The primal Lagrangians Eqs. (\ref{Solution16}, \ref{Solution20}, \ref{Solution23}) elucidate how sensitive the dynamics of a system depends on the hierarchy of immersions and actions of several environments. Just for completeness, we illustrate the case, where the system Eq. (\ref{Solution23}) is immersed into an environment (III), described by the function $b[q(t),\,t]$. The set-relation then reads
\begin{equation}
\left\{\left\{\{\rm{free\,particles}\,\subset\,\rm{environment (I)}\}\,\leftarrow\,\rm{harmonic\,trap}\right\}\,\leftarrow\,\rm{environment (II)}\right\}\,\subset\,\rm{environment\,(III)}\quad,
\label{Solution24}\end{equation}
and the primal Lagrangian becomes
\begin{equation}
\mathcal{L}(\dot{q},\,q)=b^{-2}[q(t),\,t]\left(\left\{\left\{\left\{\exp[-\,\mu_{1}\,t]\frac{m}{2}\,\dot{q}(t)^{2}\right\}-U_{\rm{E}}[q(t)]\right\}+\sigma_{2}\,q(t)\exp\left[\int_{t_{0}}^{t}d\tau\,\partial_{q}^{2}U_{\rm{E}}[q(\tau)]\right]\right\}\right)-\mathcal{V}[q(t),\,t]\,,
\label{Solution25}\end{equation}
where the potential $\mathcal{V}[q(t),\,t]$ subsumes all additional potentials, that depend on $b[q(t),\,t]$, see \cite{Jurisch1}.

By our discussion we understand that immersions and actions of environments mostly can be implemented directly into the Lagrangian without complications. Only the action of an additional stochastic process on the Newtonian equation of motion requires to go through the formalism we have developed above.

\section{Remark}
As a last point, we shall remark that in exotic field-theory methods have been developed, which, under certain circumstances, can remove the Ostrogradsky-instability from the Hamiltonian, see e.g. Chen et. al. \cite{Chen}. The method of choice is the Dirac-constraint \cite{Dirac}. If the Ostrogradsky-instability is removable, this leads to a reduction of the phase-space, where the system then evolves on a restricted hyper-surface only. Chen et. al. \cite{Chen} successfully probe the method of Dirac-constraints on a variety of elementary problems, including the Pais-Uhlenbeck oscillator \cite{Pais}. However, if a reduction of the phase-space is possible, then this comes at the cost that the resulting Lagrangian is completely remote from the initial problem. Furthermore, in order to be successful, artificial auxiliary terms must be added to the initial ill-defined Lagrangian by Lagrange-multipliers, which are constructed in a way that the Dirac-constraints can act as wished. From a sane bottom-up view, however, such approaches are more than questionable. To enforce something by tinkering, which is just not there ignores a distinct caveat.

Our suggestions is built upon the contrary, in that we enlarge the dimension of the phase-space. This allows us not only to resolve the second order derivative of the Newtonian equation of motion, but also the inclusion of the stochastic process by applying the Onsager-Machlup theory straight.

\section{Conclusion}
By using Ostrogradsky's theorem, we ruled out the possibility to reinterpret Newton's equation of motion with Stokes-friction and stochastic noise as a Langevin-equation, since the corresponding Onsager-Machlup Lagrangian contains second order derivatives in a non-degenerate way. Such a Lagrangian is ill-defined and there is no physics in it. This shows that Newton's equation of motion and the Langevin-equation are something completely different. 

Furthermore, we suggested a method of how to treat a system where Newton's equation of motion is distorted by noise. Our ansatz is built upon the canonical formalism, and the results we achieved seem to be sound. We were able to derive an Onsager-Machlup Lagrangian, that describes the effects of the stochastic distortion by a force, which additionally influences of the most probable path a particle takes through a disordered environment.

Last, we gave arguments of how to understand the hierarchy of environments, that prove useful for the analysis of systems of higher complexity. This has shown that the structure of the primal Lagrangian sensitively depends on the hierarchy of immersions and actions of the environments.

\end{document}